\documentclass[conference]{IEEEtran}
\usepackage{amsmath}
\usepackage{amssymb}
\usepackage{graphicx}
\usepackage{subcaption}
\usepackage{booktabs}    
\usepackage{siunitx}
\usepackage{tabularx,multirow}
\usepackage{mathtools}

\ifCLASSINFOpdf
\else
\fi
\begin{document}
%
\title{Use ADAS Data to Predict Near-Miss Events: A Group-Based Zero-Inflated Poisson Approach}
%
%
%


\author{
\IEEEauthorblockN{%
Xinbo Zhang\textsuperscript{1},
Montserrat Guillen\textsuperscript{2},
Lishuai Li\textsuperscript{3},
Xin Li\textsuperscript{4},
Frank Youhua Chen\textsuperscript{1*}}
\IEEEauthorblockA{%
\textsuperscript{1}Department of Decision Analytics and Operations, City University of Hong Kong, Hong Kong SAR, China \\
\textsuperscript{2}Department of Econometrics, University of Barcelona, Barcelona, Spain \\
\textsuperscript{3}Department of Data Science, City University of Hong Kong, Hong Kong SAR, China \\
\textsuperscript{4}Department of Information Systems, City University of Hong Kong, Hong Kong SAR, China}
\\
Email: xinbo.zhang@cityu.edu.hk, mguillen@ub.edu, lishuai.li@cityu.edu.hk, Xin.Li@cityu.edu.hk, youhchen@cityu.edu.hk
}

\maketitle

\begin{abstract}
Driving behavior big data leverages multi-sensor telematics to understand how people drive and powers applications such as risk evaluation, insurance pricing, and targeted intervention. Usage-based insurance (UBI) built on these data has become mainstream. Telematics‐captured near-miss events (NMEs) provide a timely alternative to claim-based risk, but weekly NMEs are sparse, highly zero-inflated, and behaviorally heterogeneous even after exposure normalization. Analyzing multi-sensor telematics and ADAS warnings, we show that the traditional statistical models underfit the dataset. We address these challenges by proposing a set of zero-inflated Poisson (ZIP) frameworks that learn latent behavior groups and fits offset-based count models via EM to yield calibrated, interpretable weekly risk predictions. Using a naturalistic dataset from a fleet of 354 commercial drivers over a year, during which the drivers completed 287,511 trips and logged 8,142,896 km in total, our results show consistent improvements over baselines and prior telematics models, with lower AIC/BIC values in-sample and better calibration out-of-sample. We also conducted sensitivity analyses on the EM-based grouping for the number of clusters, finding that the gains were robust and interpretable. Practically, this supports context-aware ratemaking on a weekly basis and fairer premiums by recognizing heterogeneous driving styles.
\end{abstract}


%
\IEEEpeerreviewmaketitle

\section{Introduction}
\label{sec:intro}

Driving behavior big data aims at understanding drivers through various telematics methods. The analysis of those data has wide application in various domains, such as evaluating driving risks, deciding insurance premiums, and intervening in driving misbehavior. For example, in recent decades, big data-based usage-based insurance (UBI) became popular. According to \cite{MM2025usagebased}, the UBI market is projected to grow from USD 43.38 billion in 2023 to USD 70.46 billion by 2030.

In this paper, we are interested in studying telematics data for near-miss event (NME) prediction. NMEs refer to sudden activities in driving, including acceleration, braking intensity, turning radius, etc. The collation and analysis of NME enables the government to develop road safety standards for automobiles in various conditions and in specific traffic situations. From an industry perspective, NME enables a structural shift from low-frequency, claim-based rating to high-frequency, behavior-aware pricing, accelerating the transition from Pay-As-You-Drive (PAYD) to Pay-How-You-Drive (PHYD) products. Operationally, this shifts focus on safety management from `Post-event Claims' to `Pre-event Training'. NMEs can also replace crashes as a risk indicator for risk modeling. Compared with low-severity incident data, the more frequent NMEs provide the possibility of richer behavioral insights.

In the big data literature on driving behavior, researchers often focus on driving mobility factors such as time, mileage, and speed, using personal driving information to predict driving risks and improve insurance pricing \cite{jun2011differences, chen2019effects, ho2022all}. Most studies leverage On-Board Diagnostic (OBD) and Global Positioning System (GPS) trajectory data to extract driving behavior features from basic signals. e.g., instant speed, acceleration/braking intensity, frequency of sudden acceleration/braking events, lane departure events, driving time distribution, and road type proportion, etc. The automotive industry has actively addressed driving safety concerns through the implementation of advanced driver assistance systems (ADAS), harnessing Internet of Things (IoT) technologies to automate, optimize, and improve vehicle functions. ADAS systems deliver critical risk-related feedback, alerting drivers to potential threats and, consequently, supporting safer decision-making. \cite{yang2025general, antony2021advanced, eby2018prevalence}. Currently, very few studies have explored driving behavior learning by fully assessing multi‐sensor telematics to characterize near‐miss events and deliver targeted feedback. 

In driving behavior studies, certain data features pose persistent challenges. Most automobile insurance databases record a large number of policyholders with zero claims, and zero inflation may be exacerbated by reporting incentives (e.g., deductibles or bonus-malus penalties) \cite{chiappori2000testing, dionne1992automobile}. In contrast, when NMEs are captured in real-time via telematics, all events occurring while the device is active are observed. Nevertheless, zero counts remain frequent at the driver-week level because NMEs are rare within short aggregation windows and drivers differ in exposure. As a result, a standard Poisson model tends to underestimate the probability of zero, motivating zero-inflated specifications and, when necessary, overdispersion-robust variants. Such datasets typically exhibit a long-tailed distribution, characterized by a high concentration of zero counts alongside a sparse but highly uneven spread of non-zero counts. In statistical terms, a long tail means that, beyond the dominant zeros, most non-zero observations are very small, while a tiny fraction of cases may record very high counts, stretching the distribution's tail. This pattern causes problems when fitting a standard model; the probability of zero events is often underestimated, so adjustments are needed to account for the higher-than-expected zero count. The zero-inflated Poisson (ZIP) model is a natural choice for handling `excess zeros' in claim or near-miss data, which we take to tackle the problem. 

Furthermore, accurately characterizing driver behavior is essential in driving risk analysis. Different types of vehicles and drivers show distinct patterns: sports car drivers often perform high-risk maneuvers such as rapid acceleration and sharp turns, while conservative drivers favor gentle acceleration and gradual braking. Grouping data before modeling also helps address two common issues: data sparsity and data inconsistency. To overcome these challenges, we propose grouped predictive analytics to group drivers in clusters by style or car model and then fit separate models for each group.

Specifically, in this paper, we frame the problem of NME prediction as a time series forecasting problem (on a weekly basis). We propose a Grouped ZIP (G-ZIP) model to handle the “excess of zeros” in the weekly near-miss counts. The model gathers drivers into behaviorally homogeneous groups (e.g., by car model or driving style) and fits a separate ZIP model to each cluster to capture heterogeneity. To address the long-tail distribution of non-zero counts, we develop a Grouped Zero-Inflated Generalized Poisson (G-ZIGP) model that extends G-ZIP to accommodate both excess zeros and overdispersion. Furthermore, we analyze a historical sensor stream ADAS dataset, and a driver profiles UBI dataset to compute personalized premiums, updating rates weekly based on predicted near-miss risk.

We conduct experiments on the ADAS-GNCC telematics dataset of 354 drivers in Ireland from April 2021 to March 2022, aggregated to 12,528 driver-weeks with GNSS traces, containing warning-based ADAS event information and 16 contextual attributes \cite{masello2024dataset}. Our performance achieves improvements over strong baselines in AIC/BIC and RMSE and remains robust under sensitivity analysis, where the number of behavior groups is varied. In conclusion, our approach not only advances automobile telematics risk modeling but also provides actionable insights for insurers and managers through an end-to-end prediction and optimization pipeline.

The paper is organized as follows. Section 2 introduces the theoretical foundation and development of related work. Section 3 proposes our main behavior prediction model. Section 4 conducts experiments, and Section 5 concludes the work.

\section{Related Work}
\label{sec:rela}

\subsection{Driver Behavior Modeling} 

Driver behavior has received considerable attention over recent decades, as extensively reviewed in \cite{kaplan2015driver, meiring2015review}. Researchers have increasingly focused on diverse sensor data and semantic features extracted from various data sources to enhance risk assessments. Early studies predominantly relied on GPS tracking data obtained from vehicles \cite{zheng2015trajectory}, smartphone-embedded sensor readings \cite{castignani2015driver}, and video footage from in-vehicle cameras \cite{tran2012modeling}. Masello et al. \cite{masello2025predictive} utilized advanced vehicle technologies such as ADAS along with GNSS-based geolocation data to refine driver risk prediction.  Thanks to recent advancements and developments in data acquisition techniques, richer datasets are emerging from multiple channels, providing a more comprehensive understanding of driver behavior. For instance, He et al. \cite{he2018profiling} integrated UBI and OBD data to develop a trajectory-based driver profiling method, which aimed at extracting risk-related behavioral patterns. Subsequently, the trajectory-based method was refined to employ OBD data for behavior analysis and risk prediction \cite{he2018pbe}. Xie et al. \cite{xie2017analysis} explored the complexity of driving behaviors through the fusion of offline GPS trackers and OBD information. Moreover, Ho et al. \cite{ho2022all} extended behavioral profiling further by identifying semantic driving features from real-time streams of GPS, OBD, and in-vehicle camera (IVC) data, considering both individual trip characteristics and overall driver-level patterns. 

Beyond telematics and driver-specific information, external contextual variables such as weather conditions have also been increasingly incorporated into risk assessments. Mornet et al. \cite{mornet2015index} constructed an economic index for insurance risk management based on historical wind speed records in France. Similarly, Gao and Shi \cite{gao2022leveraging} quantified the impact of hailstorms on insurance claims using real insurer data from the United States. More recently, Reig Torra et al. \cite{reig2023weather} developed claim frequency models by integrating telematics and detailed weather data into frameworks. 

Most studies take a feature-based approach, extracting driving behavior features from basic signals. The use of statistical techniques and machine learning algorithms has been well studied. Guillen et al. \cite{guillen2019use} include distance travelled per year as part of an offset in a zero-inflated Poisson model to predict the excess of zeros. Then, they used negative binomial (NB) regression to model the number of near-miss events \cite{guillen2020can}. Yanez et al. \cite{yanez2025weekly} refer to the bonus-malus credibility models proposed by Lemaire et al. \cite{lemaire2013automobile} and redefine the model within the generalized linear models (GLM) framework. Ho et al. \cite{ho2022all} proposed mobility-based risk assessment (MRA) as a generalized UBI solution, implementing a Classification and Regression Tree (CART) algorithm with Gini impurity calculation \cite{loh2011classification} to produce risk probabilities that can readily be integrated into the ORC model. He et al. \cite{he2018pbe} employed the popular Gradient Boosting Decision Tree (GBDT) as a multi-class classifier to formulate a driver behavior model. These studies are then applied to associate these features with historical accident and claim records. The resulting analyses support the construction of driving score systems or driving risk evaluation frameworks \cite{jun2007relationships, xie2017analysis, he2018profiling}. Based on the risk evaluation score, insurance companies have the opportunity to adjust premiums dynamically, transforming `average pricing' to `individual pricing'. However, these methods ignore the real-time risk warning and behavior intervention mechanisms. This leaves ample scope for the development of models and applications tailored to each driver's characteristic behavior in the future.

\subsection{Applications from Driving Behavior Modeling} 

The introduction of UBI models, such as PAYD \cite{paefgen2014multivariate} and subsequent PHYD schemes, represents a notable advance toward greater pricing flexibility supported by driving behavior modeling. Similarly, He et al. \cite{he2018profiling} integrated both mileage-based metrics and trajectory-level behavioral insights into their proposed dynamic pricing model. Yanez et al. \cite{yanez2025weekly} proposed adapting traditional Bonus-Malus Systems (BMS) for telematics-enabled claim frequency prediction, enhancing dynamic pricing effectiveness. Driving behavior modeling can also be used in risk management. Guillen et al. \cite{guillen2021near} introduced a near-miss event frequency model specifically designed to capture risk indicators from driving data. Zhu et al. \cite{zhu2017bayesian} conducted a study on the driving context, which has been shown to improve the performance of risk assessment models. The authors propose a Bayesian Network model to investigate the relationship between driving behavior and risk assessment. Wang et al. \cite{wang2015driving} integrated driving behavior, vehicle features and contextual variables for a new risk assessment CART method based on near-misses. Masello et. al. \cite{masello2023using} applied Shapley Additive Explanations (SHAP), which was employed from the perspective of risk assessment, to conduct a comprehensive analysis of near-misses and a series of contextual driving attributes. More recently, Masello et al. \cite{masello2025predictive} considered dual-model frameworks by separately modeling claim frequency and claim occurrence probabilities, thereby computing individualized premiums that better reflect driver-specific risks and actual driving behaviors.

\subsection{Research Gaps and Our Contributions}

Despite notable progress in driver behavior modeling and its applications, the extant literature still exhibits several limitations:

\begin{enumerate}
  \item \textbf{Zero inflation and long-tail distribution.} Zero counts remain frequent at the driver-week level of NMEs. Beyond the dominant zeros, most non-zero counts are small while a few cases are very large, yielding long-tailed, overdispersed outcomes. Conventional existing models yield mis-specified likelihoods and biased uncertainty, weakening the performance.
  
  \item \textbf{Heterogeneity driving behavior data.} Many studies inconsistently adjust for exposure intensity, such as mileage, and for contextual data, such as road condition or weather condition. Integration of ADAS warning systems with GNSS contextual data within unified frameworks remains limited.
  
  \item \textbf{Data sparsity and data availability.} Privacy often leads to having only short observation windows for drivers; also, drivers may be unwilling to cooperate or may share short-period data. The lack of time stream data makes it a significant challenge in the development of a reliable risk prediction model for each individual. Meanwhile, aggregating all drivers to fit a single model can introduce inconsistency because different drivers often have different driving behaviors.
  
\end{enumerate}

Close to our heterogeneity treatment is the mixture-of-experts with random effects for a-posteriori ratemaking \cite{tseung2023improving}. Unlike their policy-year, claim-based setting, building upon Masello et al. \cite{masello2025predictive} and the dataset \cite{masello2024dataset}, we operate at the weekly telematics resolution on near-miss counts and address excess zeros and heavy tails via an exposure-adjusted set of zero-inflated models with EM-based latent grouping. The proposed approach captures excess zeros and long-tails while accommodating heterogeneity, and it yields deployable risk scores for risk assessment. Experiments show that our method outperforms the existing model and leads to more stable portfolio and premium metrics.

\section{Problem Setup}
\label{sec:prob}

This section presents the dataset used for NME risk prediction, defines our main objectives and the NME construct, and formally states the prediction problem.

\subsection{Data}
\label{subsec:data}

\begin{table*}[t]
\centering
\caption{Data Description}
\label{tab:mendelydata}
\footnotesize
\begin{tabularx}{\textwidth}{l l l X}
\toprule
\textbf{Category} & \textbf{Attribute} & \textbf{Values} & \textbf{Description} \\
\midrule
\multirow{17}{*}{Driving context}
  & mean\_speed\_limit [km/h] & $(0,120]$ & Mean legal speed limit along driven segments. \\
  & mean\_weather\_temperature [$^\circ$C] & $[0,25]$ & Mean ambient temperature during driving sessions. \\
  & mean\_weather\_visibility [m] & $(0,10{,}000]$ & Mean visibility during driving sessions. \\
  & mean\_weather\_wind\_speed [km/h] & $[0,46]$ & Mean wind speed during driving sessions. \\
  & prop\_clear\_weather & $[0,1]$ & Share of exposure with clear weather. \\
  & prop\_congested & $[0,1]$ & Share of samples with avg.\ traffic speed $< 25\%$ of the limit. \\
  & prop\_more\_than\_one\_lane & $[0,1]$ & Share of road segments with $\ge$2 lanes. \\
  & prop\_motorway & $[0,1]$ & Share of exposure recorded on motorway. \\
  & prop\_road\_quality\_moderate & $[0,1]$ & Share with moderate pavement quality (e.g., IRI $\le 6$). \\
  & prop\_rural & $[0,1]$ & Share on rural routes (urban $\approx 1-\,$prop\_rural$-\,$prop\_motorway). \\
  & prop\_slope\_flat & $[0,1]$ & Share of road slope in $[-2^\circ,2^\circ]$. \\
  & sum\_animal\_crossing\_sign & $[0,41]$ & Count of animal‐crossing signs passed. \\
  & sum\_pedestrian\_crossing\_sign & $[0,51]$ & Count of pedestrian/zebra crossings passed. \\
  & sum\_roundabout & $[0,559]$ & Count of roundabouts encountered. \\
  & sum\_stop\_sign & $[0,48]$ & Count of stop signs passed. \\
  & sum\_traffic\_light & $[0,257]$ & Count of traffic lights passed. \\
  & sum\_yield\_sign & $[0,148]$ & Count of yield (give‐way) signs passed. \\
\midrule
\multirow{4}{*}{Driving behavior}
  & sum\_harsh\_acceleration & $[0,3996]$ & Longitudinal acceleration events $> 6\,\mathrm{m/s^2}$. \\
  & sum\_harsh\_braking & $[0,2040]$ & Longitudinal deceleration events $> 6\,\mathrm{m/s^2}$. \\
  & sum\_speeding\_serious & $[0,300]$ & Speeding $\ge 20$ km/h above the legal limit (weekly count). \\
\midrule
\multirow{7}{*}{ADAS warnings}
  & sum\_fatigue\_driving & $[0,758]$ & Fatigue/drowsiness warning detected by driver monitoring. \\
  & sum\_forward\_collision & $[0,141]$ & Potential collision warning (e.g., closing on a stopped car). \\
  & sum\_driver\_inattention & $[0,1603]$ & Inattention/distracted driving (e.g., gaze off road). \\
  & sum\_driver\_smoking & $[0,736]$ & Smoking while driving detected. \\
  & sum\_driver\_making\_calls & $[0,268]$ & Phone call while driving detected. \\
  & sum\_lane\_departure & $[0,1374]$ & Lane departure or lane change without indicators. \\
  & sum\_too\_close\_distance & $[0,573]$ & Following distance too short at speed $>30$ km/h. \\
\midrule
Driving exposure
  & total\_distance [km] & $[10,3423]$ & Total weekly driven distance. \\
\midrule
Vehicle information
  & engine\_capacity [thousands cc] & $[1.5,2.3]$ & Engine displacement (thousands of cubic centimeters). \\
\midrule
\multirow{2}{*}{Claim information}
  & exposure\_in\_weeks & $\mathbb{N}_{\ge 1}$ & Observation/contract weeks used as exposure (offset). \\
  & claims\_count & $[0,2]$ & Number of at‐fault claims in the history window. \\
\bottomrule
\end{tabularx}
\vspace{2pt}
\end{table*}

We analyze a naturalistic fleet dataset covering 354 drivers operating in Ireland from April 2021 to March 2022. Across the campaign, drivers completed 287{,}511 trips and logged 8{,}142{,}896 km in total. On average, a driver undertook about five trips per day and covered roughly 143 km, with an average monitoring duration of 277 days. The fleet was monitored with telematics tracking devices and warning-based ADAS that triggered alarms about distraction-related events. With this information, all drivers received feedback about their driving patterns and attended quarterly coaching sessions to meet the fleet’s standards regarding road safety.

The data contains two parts: warning-based ADAS and contextual GNCC data. All signals are timestamped and aligned, allowing each ADAS event to be matched to the route being driven and its surroundings. The GNSS traces are then enriched with 16 context variables that describe the road environment, traffic conditions, road signs, weather, etc.. Driver behavior is captured from two sources: (i) telematics anomalies—events that indicate risky vehicle dynamics, such as harsh acceleration, harsh braking, and speeding; and (ii) camera-based ADAS warnings triggered when specific conditions are met. The ADAS set includes phone calls, smoking, fatigue, lane departure, etc.. To build risk profiles, we aggregate all variables in weekly windows, resulting in 12{,}528 driver-weeks with the attributes listed in Table \ref{tab:mendelydata}. The dataset is publicly available in \cite{masello2024dataset}.

\subsection{Near‐Miss Events}
\label{subsec:nme}

We define NMEs as safety–critical incidents detected either by vehicle kinematics anomalies or by on–board ADAS warnings that indicate an immediate risk. In our data, the NME set $\mathcal{E}$ is defined as:

\[
\mathcal{E}=\left\{
\begin{array}{@{}l l@{}}
\text{harsh\_braking},     & \text{harsh\_acceleration},\\
\text{serious\_speeding},  & \text{forward\_collision},\\
\text{lane\_departure},    & \text{too\_close\_distance}
\end{array}
\right\}.
\]

Let $C^{(e)}_{i,t}$ be the weekly count for event type $e\in\mathcal{E}$ for driver $i$ in week $t$. Apart from analyzing NMEs individually, we also aggregate to a combination NME count:
\[
N_{i,t} \;=\; \sum_{e\in\mathcal{E}} C^{(e)}_{i,t},
\]

\subsection{Problem Description}
\label{subsec:probdes}

We consider a set of drivers indexed by $i\in\{1,\dots,I\}$ observed over weeks $t=1,\dots,T_i$. Let $\mathcal{D}=\{(i,t): i=1,\dots,I;\ t=1,\dots,T_i\}$ denote the set of all driver--week observations and let $n=\lvert\mathcal{D}\rvert$ be the total number of observations.
For each $(i,t)\in\mathcal{D}$, let $C^{(e)}_{i,t}$ be the weekly count of NME type $e\in\mathcal{E}$, the combination NME count $N_{i,t}$ is defined above. Let $\mathbf{x}_{i,t}\in\mathbb{R}^k$ be a vector of $k$ attributes, and let $\mathbf{x}_{i,t}=(x_{i,t,1},\dots,x_{i,t,k})^\top$. Our problem is to model individual $C^{(e)}_{i,t}$ and combination \(N_{i,t}\) directly with the proposed set of models. The models are interpreted using GLM coefficients for NME frequency modeling.

\section{Model Specification}
\label{sec:modspe}

This section presents the risk-assessment methodology for predicting NMEs. Leveraging warning-based ADAS signals enriched with GNSS-derived contextual features, we build a modeling pipeline that addresses the characteristic challenges of telematics data—excess zeros, longs, driver heterogeneity, sparsity, and inconsistency, which undermine a plain Poisson baseline. We introduce (i) a zero-inflated Poisson (ZIP), (ii) a group-based ZIP to capture between-driver heterogeneity (G-ZIP), and (iii) a group-based zero-inflated generalized Poisson (G-ZIGP) to jointly accommodate zero inflation and dispersion. Interpretability is maintained through a GLM formulation, reporting coefficients as log-rate effects on weekly NME frequency.

\subsection{Poisson Model}
\label{subsec:poi}

GLMs are used to model the relationships between the number of NMEs in a given period and driver profile attributes, assuming that the number of NMEs follows a Poisson distribution. This method follows the methodology positioned by Gullien et al. \cite{guillen2021near}.

We model both the type-specific counts $C^{(e)}_{i,t}$ for each $e\in\mathcal{E}$ and the combination NME $N_{i,t}=\sum_{e\in\mathcal{E}} C^{(e)}_{i,t}$. For each driver–week $(i,t)$ we observe covariates $\mathbf{x}_{i,t}\in\mathbb{R}^k$ measured by the weekly total distance $E_{i,t}>0$. Here, $\lambda^{(e)}{i,t}$ denotes the expected rate per unit exposure of NME type $e$ in week $t$ for driver $i$, and the aggregate rate $\Lambda_{i,t}$ is the expected total NME rate per unit exposure in week $t$. Rates are modeled with GLMs using log links as follows.

\begin{equation}
C^{(e)}_{i,t}\mid \mathbf{x}_{i,t},E_{i,t} \;\sim\; 
\mathrm{Poisson}\!\big(E_{i,t}\,\lambda^{(e)}_{i,t}\big),
\end{equation}

\begin{equation}
\log \lambda^{(e)}_{i,t}
\;=\; \alpha^{(e)}_0 + \mathbf{x}_{i,t}^{\!\top}\boldsymbol{\beta}^{(e)},
\quad e\in\mathcal{E}.
\end{equation}

For combination NME modeling, we have 

\begin{equation}
N_{i,t}\mid \mathbf{x}_{i,t},E_{i,t} \;\sim\; 
\mathrm{Poisson}\!\big(E_{i,t}\,\Lambda_{i,t}\big),
\end{equation}

\begin{equation}
\log \Lambda_{i,t}= \alpha_0 + \mathbf{x}_{i,t}^{\!\top}\boldsymbol{\beta}.
\end{equation}

\subsection{Zero-Inflated Poisson Model}
\label{subsec:zip}

The ZIP regression is a model for count data with an excess of zeros. In the ZIP model, $\pi_{i,t}$ is the probability of the structural zero state, and $(1-\pi_i)$ the probability of the complementary state. The complementary state follows a Poisson law with the same exposure $E_{i,t}$ as in Section~\ref{subsec:poi}. Let $\gamma_0\in\mathbb{R}$ be the intercept of the zero–inflation model, and let $\boldsymbol{\gamma}\in\mathbb{R}^{k}$ be the coefficient vector on the covariates $\mathbf{x}_{i,t}\in\mathbb{R}^{k}$, so that $\pi_{i,t}\in(0,1)$ is ensured by the logit link. We specify the link function as:

\begin{equation}
    \mathrm{logit}\,\pi_{i,t} = \gamma_0 + \mathbf{x}_{i,t}^{\!\top}\boldsymbol{\gamma},
\end{equation}

\begin{equation}
    \log \lambda^{(e)}_{i,t} = \alpha^{(e)}_0 + \mathbf{x}_{i,t}^{\!\top}\boldsymbol{\beta}^{(e)}, \quad e\in\mathcal{E}.
\end{equation}

For each type $e\in\mathcal{E}$, let the complementary state follow a Poisson law. The probability mass function of ZIP is:

\begin{multline}
\Pr(C^{(e)}_{i,t}=0 \mid \mathbf{x}_{i,t},E_{i,t})
= \pi_{i,t} \\
+ (1-\pi_{i,t})\,\exp(-E_{i,t}\lambda^{(e)}_{i,t}).
\label{eq:zip-pmf1}
\end{multline}

When the count of individual NME $k\geq1$, we have:

\begin{multline}
\Pr(C^{(e)}_{i,t}=k \mid \mathbf{x}_{i,t},E_{i,t})
= (1-\pi_{i,t}) \\
\times \,\exp(-E_{i,t}\lambda^{(e)}_{i,t})\,
   \frac{(E_{i,t}\lambda^{(e)}_{i,t})^k}{k!}, \quad k\in\mathbb{N}_{+}.
\label{eq:zip-pmf2}
\end{multline}

Similarly, for the combination NME, we have:

\begin{multline}
\Pr(N_{i,t}=0 \mid \mathbf{x}_{i,t},E_{i,t}) = \pi_{i,t} \\ + (1-\pi_{i,t})\,\exp(-E_{i,t}\Lambda_{i,t}),
\label{eq:zip-pmf3}
\end{multline}

When the count of combination NME $k\geq1$, we have:

\begin{multline}
\Pr(N_{i,t}=k \mid \mathbf{x}_{i,t},E_{i,t}) = (1-\pi_{i,t}) \\
\times \,\exp(-E_{i,t}\Lambda_{i,t})\,
\frac{(E_{i,t}\Lambda_{i,t})^k}{k!}, \quad k\in\mathbb{N}_{+},
\label{eq:zip-pmf4}
\end{multline}

\subsection{Group‐Based Zero‐Inflated Poisson Model}
\label{sec:zip_model}

Given the weekly driver NME dataset, it is not straightforward to predict a driver’s future risk by simply applying the ZIP model due to data sparsity and inconsistency. To address the issue of heterogeneity, we improved the ZIP model and proposed a group-based ZIP model. Drivers' entire driving behavior can be regarded as several patterns based on environment, driving style, etc. Within the same group, drivers share similar driving behavior. We can use the data within the same group to train one ZIP model, overcoming the sparsity issue. Training data within different groups increases the effective sample size and stabilizes estimation, while allowing parameters to vary across groups mitigates inconsistency and improves short-horizon risk forecasts. In practice, groups can be obtained via latent-class Expectation-Maximization (EM) clustering; the resulting ensemble of ZIP models replaces a one-size-fits-all specification and better reflects the diversity of driving patterns.

We partition drivers into $G$ behaviorally homogeneous groups and fit a group-specific ZIP with the same exposure offset $\log E_{i,t}$ used in Section~\ref{subsec:poi}. Within each group the ZIP is based on Section~\ref{subsec:zip}. Let the latent membership be $Z_i\in\{1,\dots,G\}$ with mixing weights $\omega_g=\Pr(Z_i=g)$, we therefore have $\sum_{g=1}^G\omega_g=1$. Within group $g$, we use
\begin{equation}
\mathrm{logit}\,\pi_{i,t}^{(g)} = \gamma^{(g)}_0 + \mathbf{x}_{i,t}^{\!\top}\boldsymbol{\gamma}^{(g)},
\end{equation}

\begin{equation}
\log \Lambda_{i,t}^{(g)} = \alpha^{(g)}_0 + \mathbf{x}_{i,t}^{\!\top}\boldsymbol{\beta}^{(g)},  
\end{equation}

Conditional on $Z_i=g$, the probability mass function is similar to Equation \eqref{eq:zip-pmf1} to \eqref{eq:zip-pmf4} with the substitutions $(\pi_{i,t},m_{i,t})$ with $(\pi^{(g)}_{i,t},m^{(g)}_{i,t})$ (or $m^{(e,g)}_{i,t}$ for type $e$). For simplicity, we present only the ZIP probability mass function for the individual NME counts:

\begin{multline}
\Pr(C_{i,t}^{(e)}=0 \mid Z_i=g,\mathbf{x}_{i,t},E_{i,t})
= \pi_{i,t}^{(g)} \\ + (1-\pi_{i,t}^{(g)})\exp(-E_{i,t}\lambda_{i,t}^{(e,g)}),
\label{eq:gzip-pmf1}
\end{multline}

\begin{multline}
\Pr(C_{i,t}^{(e)}=k \mid Z_i=g,\mathbf{x}_{i,t},E_{i,t})
= (1-\pi_{i,t}^{(g)})\\ \times\exp(-E_{i,t}\lambda_{i,t}^{(e,g)})\,
   \frac{(E_{i,t}\lambda_{i,t}^{(e,g)})^k}{k!},\;\; k\in\mathbb{N}_+.  
\label{eq:gzip-pmf2}
\end{multline}

The observed distribution is a finite mixture of the group-specific ZIPs. The corresponding marginal model is $\Pr(C_{i,t}^{(e)}=0 \mid \mathbf{x}_{i,t},E_{i,t})=\sum_{g=1}^G \omega_g \Pr(C_{i,t}^{(e)}=k \mid Z_i=g,\mathbf{x}_{i,t},E_{i,t}), k\geq 0$.

\subsection{Zero-Inflated Generalized Poisson (ZIGP)}
\label{subsec:zigp}

Furthermore, we extend ZIP by replacing the complementary Poisson law with a generalized Poisson (GP), allowing a long-tail distribution. We first introduce the Generalized Poisson with dispersion $\theta$. When $\theta=0$, GP reduces to Poisson. When $\theta>0$ induces overdispersion and a heavier right tail, while $\theta<0$ induces underdispersion. The remaining notations are the same.

For $k=0,1,2,\dots$ the GP probability mass function with mean parameter $m>0$ and dispersion $\theta$ is
\begin{equation}
\Pr(Y=k)=
\frac{m\,\big(m+\theta k\big)^{k-1}\,\exp\!\big(-m-\theta k\big)}{k!}
\end{equation}

For individual NME type $e\in\mathcal{E}$, with $m^{(e)}_{i,t}=E_{i,t}\lambda^{(e)}_{i,t}$ and dispersion $\theta^{(e)}$, the ZIGP probability mass function is

\begin{multline}
\Pr(C^{(e)}_{i,t}=0 \mid \mathbf{x}_{i,t},E_{i,t})
= \pi_{i,t} + (1-\pi_{i,t})\,e^{-m^{(e)}_{i,t}} 
\end{multline}

\begin{multline}
\Pr(C^{(e)}_{i,t}=k \mid \mathbf{x}_{i,t},E_{i,t})= \\ (1-\pi_{i,t}) \frac{m^{(e)}_{i,t}(m^{(e)}_{i,t}+\theta^{(e)}k)^{k-1}\,
      e^{-m^{(e)}_{i,t}-\theta^{(e)}k}}{k!},\quad k\in\mathbb{N}_{+} 
\end{multline}

subject to $m^{(e)}_{i,t}+\theta^{(e)}k>0$ for all relevant $k$. The combination NME model follows the same form.

\subsection{Unified EM Estimation for Grouped ZIP / ZIGP}
\label{subsec:em}

We developed the EM algorithm for modeling and driver grouping. We estimate $\big\{\omega_g,\,\boldsymbol{\eta}^{(g)}\big\}_{g=1}^G$ by EM, where $\boldsymbol{\eta}^{(g)}$ collects the group-$g$ regression parameters: for the combination NME metric of G-ZIP, $\boldsymbol{\eta}^{(g)}=\{\gamma^{(g)}_0,\boldsymbol{\gamma}^{(g)}, \alpha^{(g)}_0,\boldsymbol{\beta}^{(g)}\}$; for the individual NME metric of G-ZIP, the parameters include the type of NME $e$; and for G-ZIGP model, it includes the dispersion(s) $\theta^{(g)}$ (or $\theta^{(e,g)}$). Let $Y_{i,t}$ denote the modeled count series (either $N_{i,t}$ or a chosen $C^{(e)}_{i,t}$). Given independence over $t$ conditional on parameters, the group-$g$ likelihood contribution for driver $i$ is $L_{i}^{(g)}=\prod_{t=1}^{T_0} f^{(g)}\!\big(Y_{i,t}\mid \mathbf{x}_{i,t},E_{i,t};\,\boldsymbol{\eta}^{(g)}\big), $ where $f^{(g)}$ is the ZIP/ZIGP probability mass function.

\paragraph{E-step.}
We compute posterior memberships

\begin{multline}
\tau_{i,g}
=\Pr(Z_i=g\mid \{Y_{i,t}\}_{t=1}^{T_0})
=\frac{\omega_g\,L_{i}^{(g)}}
       {\sum_{h=1}^G \omega_h\,L_{i}^{(h)}}\,,\\\quad i=1,\dots,N,\ g=1,\dots,G.
\end{multline}

\paragraph{M-step.}
We update mixing weights

\begin{equation}
\omega_g\leftarrow \frac{1}{N}\sum_{i=1}^N \tau_{i,g},
\end{equation}

and, for each $g$, maximize the weighted log-likelihood
\begin{equation}
\max_{\boldsymbol{\eta}^{(g)}}\;
\sum_{i=1}^{N}\sum_{t=1}^{T_0} \tau_{i,g}\,
\log f^{(g)}\!\big(Y_{i,t}\mid \mathbf{x}_{i,t},E_{i,t};\,\boldsymbol{\eta}^{(g)}\big). 
\end{equation}

\paragraph{Convergence.}

We denote by $\ell$ the driver-level log-likelihood of the proposed model. EM iterations stop when the observed-data log-likelihood $\ell$ increases by less than a tolerance $\varepsilon$.
\[
\big|\ell^{(t)}-\ell^{(t-1)}\big| \le \varepsilon\,(1+|\ell^{(t)}|),
\]
or when the maximum number of iterations is reached. We record $(\widehat{\omega}_g,\widehat{\boldsymbol{\eta}}^{(g)})_{g=1}^G$ and posterior group memberships $\widehat{\tau}_{i,g}$ for downstream prediction.

\section{Experiment}

This section evaluates NMEs extracted from weekly telematics records. We compare classical baselines with zero-inflated and driver-grouped extensions, and we report results on six individual NMEs (harsh braking, harsh acceleration, serious speeding, forward collision, lane departure, and too close distance) as well as their combination NME metrics. We first summarize the modelling families considered and then detail the experimental settings, including data preprocessing, feature construction, cross-validation, and evaluation metrics in Section \ref{subsec:expset}. The main comparative results are presented in Section \ref{subsec:mainres}, while several sensitivity analyses are conducted and are discussed in Section \ref{subsec:senana}.

\subsection{Experimental setup}
\label{subsec:expset}

\begin{figure*}[!t]
  \centering
  \begin{subfigure}[b]{0.3\textwidth}
    \centering
    \includegraphics[width=\textwidth]{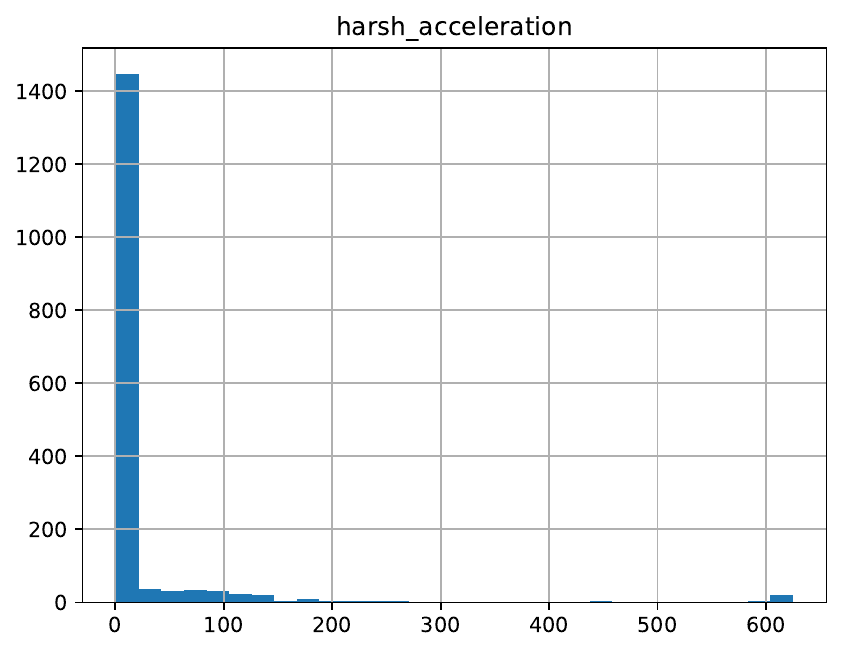}
    \caption{Harsh acceleration}
    \label{fig:hist:acceleration}
  \end{subfigure}
  \hfill
  \begin{subfigure}[b]{0.3\textwidth}
    \centering
    \includegraphics[width=\textwidth]{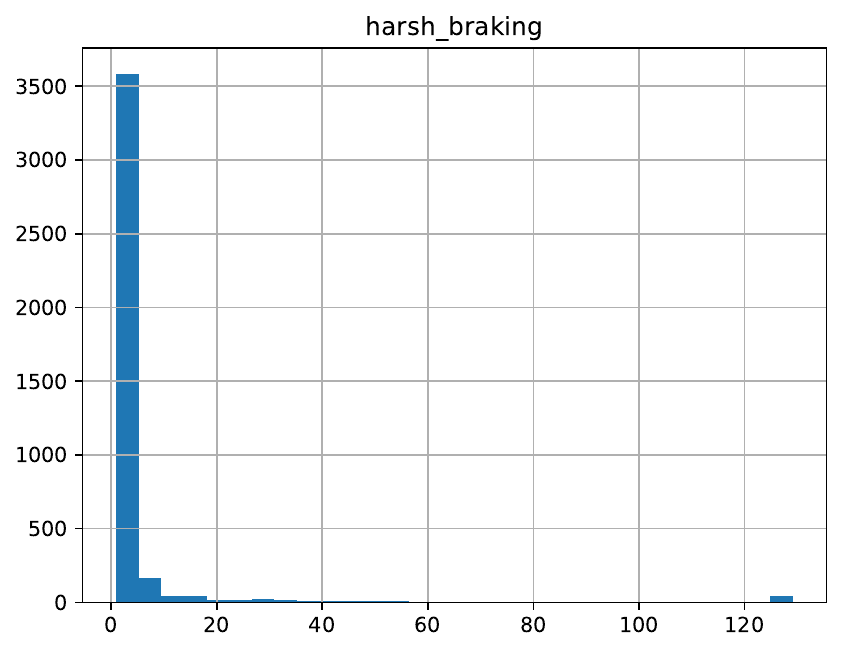}
    \caption{Harsh braking}
    \label{fig:hist:braking}
  \end{subfigure}
  \hfill
   \begin{subfigure}[b]{0.3\textwidth}
    \centering
    \includegraphics[width=\textwidth]{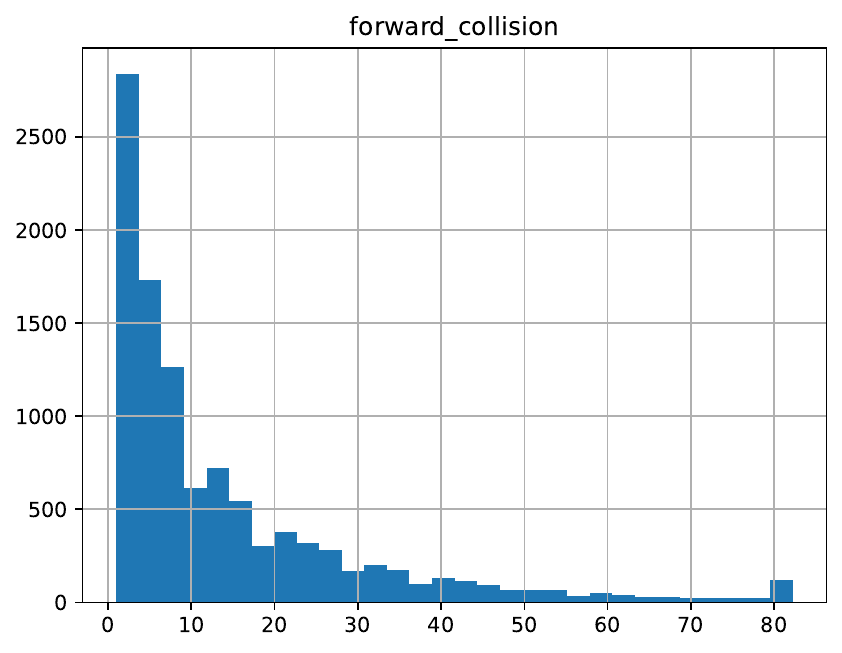}
    \caption{Forward collision}
    \label{fig:hist:forward}
  \end{subfigure}
  
  \begin{subfigure}[b]{0.3\textwidth}
    \centering
    \includegraphics[width=\textwidth]{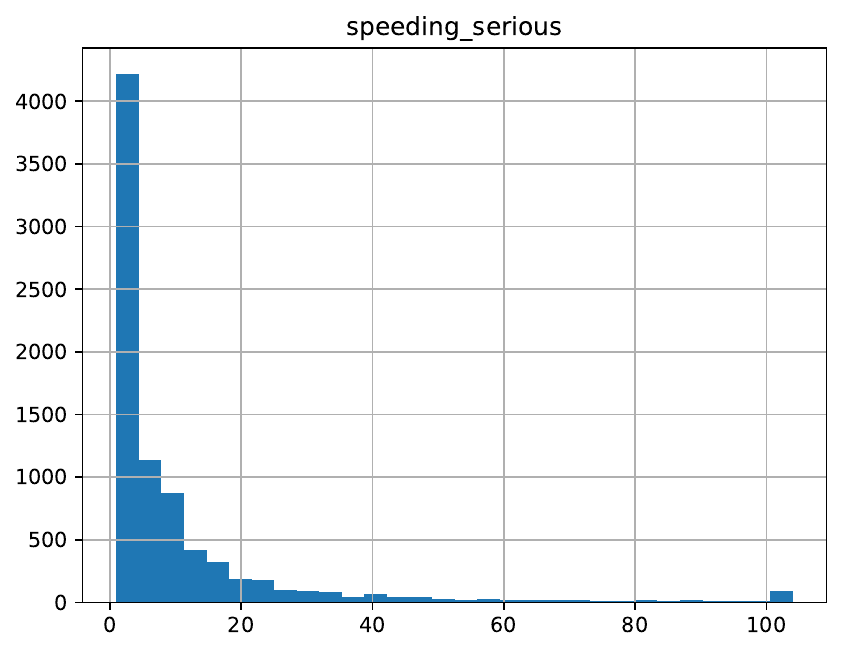}
    \caption{Serious speeding}
    \label{fig:hist:distracted}
  \end{subfigure}
  \hfill
  \begin{subfigure}[b]{0.3\textwidth}
    \centering
    \includegraphics[width=\textwidth]{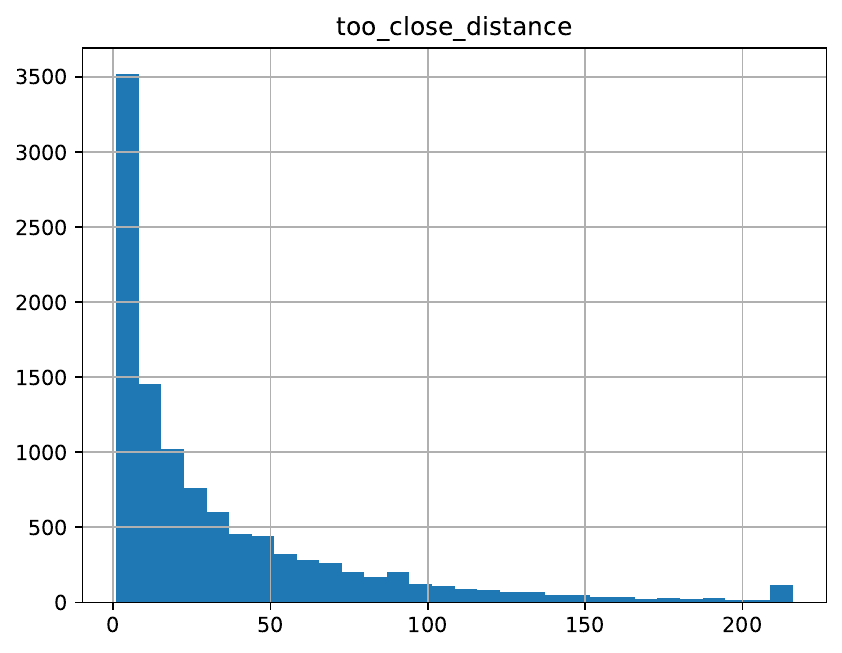}
    \caption{Too close distance}
    \label{fig:hist:distance}
  \end{subfigure}
  \hfill
  \begin{subfigure}[b]{0.3\textwidth}
    \centering
    \includegraphics[width=\textwidth]{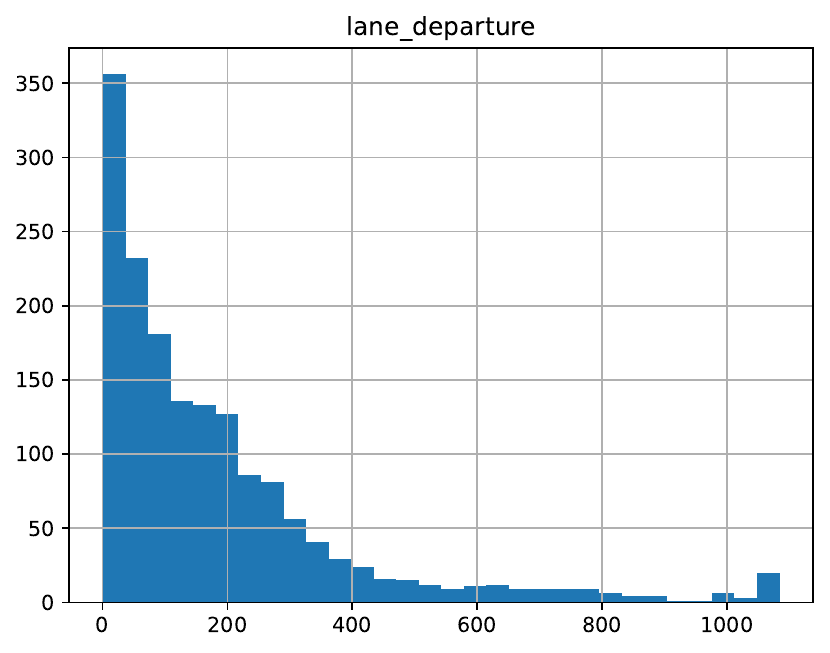}
    \caption{Lane departure}
    \label{fig:hist:lane}
  \end{subfigure}

  \caption{Non‐zero histograms for six individual NME types}
  \label{fig:nonzero‐hists}
\end{figure*}

We model six individual NMEs $C^{(e)}_{i,t}$: harsh braking, harsh acceleration, serious speeding, forward collision, lane departure, too close distance and their combination NMEs $N_{i,t}$. The exposure is the weekly total distance $E_i$; all models use the offset $\log E_i$. The NME histograms are shown in Figure \ref{fig:nonzero‐hists}, which shows excess zeros and a long-tail distribution. For the dataset, all train/test splits are performed at the driver level. We report performance under a 5-fold stratified grouped cross-validation protocol: drivers are partitioned into five non-overlapping folds while approximately preserving the proportion of drivers with at least one non-zero count across folds. Each round uses four folds for training and one for testing; metrics are averaged over folds and reported with standard deviations. We evaluate both in-sample information criteria and out-of-sample predictive accuracy:
\begin{itemize}
  \item \textbf{AIC/BIC} on the full training set for each target/model. The Akaike Information Criterion (AIC) and the Bayesian Information Criterion (BIC) are model selection criteria that balance a model's goodness of fit.
  \item \textbf{Poisson deviance} on held-out data: $D = 2\sum_{i}\left[ y_i\log\!\left(\frac{y_i}{\hat{\mu}_i}\right) - (y_i-\hat{\mu}_i)\right]$, with $y_i\log(y_i/\hat{\mu}_i)=0 \text{ when } y_i=0$.
  \item \textbf{RMSE} for counts: $\sqrt{\frac{1}{n}\sum_i (y_i-\hat{\mu}_i)^2}$.
  \item \textbf{Pearson’s $\chi^2$} goodness-of-fit: $\sum_i (y_i-\hat{\mu}_i)^2/\hat{\mu}_i$.
  \item \textbf{McFadden’s pseudo-$R^2$} relative to Possion model with offset.
  \item \textbf{Zero-event Brier score} and \textbf{zero-probability calibration}: we evaluate the predicted zero probability $\hat{p}_{0,i}$ (for ZIP/ZIGP, $\hat{p}_{0,i}=\hat{\pi}_i+(1-\hat{\pi}_i)e^{-\hat{\mu}_i}$; for Poisson, $\hat{p}_{0,i}=e^{-\hat{\mu}_i}$). The Brier score is calculated as $\frac{1}{n}\sum_i (\mathbb{1}\{y_i=0\}-\hat{p}_{0,i})^2$.
\end{itemize}

For hyperparameters, unless stated, we cap the mean-model feature count at $10$ per target after filtering, and we standardize features within the in-sample data. ZIP baselines are estimated with a maximum $800$ iterations. Let $\varepsilon=10^{-4}$, G-ZIP/G-ZIGP models are estimated with a maximum $200$ iterations per M-step, with EM convergence declared when the driver-level log-likelihood increment falls below $10^{-4}\!(1+|\ell|)$. For G-ZIGP, to ensure the dispersion parameter lies in range $\theta_g\in(-1,1)$, we use an unconstrained parameterization and estimate $a_g\in\mathbb{R}$ with $\theta_g=\tanh(a_g)$ for each group $g$; in dispersion sensitivity studies we instead fix $\theta$ on a grid and re-optimise $(\boldsymbol{\beta}_g,\gamma_g)$ only. We explore $G\in\{1,2,3,4\}$ for G-ZIP, $G\in\{1,\dots,10\}$ and generalized-Poisson dispersion $\theta\in\{-1.0,\dots, 1.0\}$ for G-ZIGP report AIC/BIC criterion.

Experiments were executed on a server with a 6-core Intel\textsuperscript{\textregistered} Xeon\textsuperscript{\textregistered} E5-2620 @ 2.00\,GHz CPU and 256\,GB RAM. Implementations use Python with statsmodels for baseline ZIP fits and custom EM solvers for grouped models.

\subsection{Main Results}
\label{subsec:mainres}

\begin{figure}[!t]
  \centering
  \includegraphics[width=\linewidth]{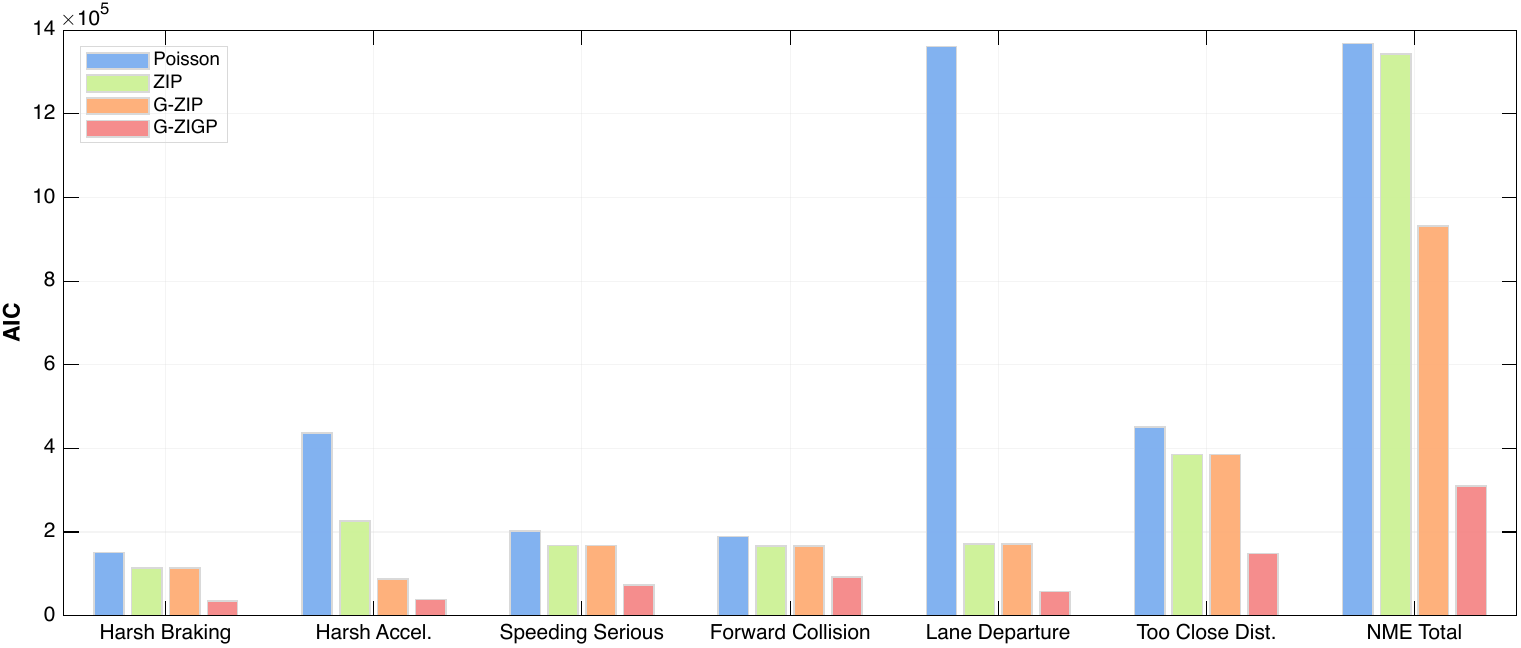}
  \caption{Telematics AIC metrics across NME categories.}
  \vspace{-0.6cm}
  \label{fig:nme-aic}
\end{figure}

Table~\ref{tab:model_eval} compares four specifications (Poisson, ZIP, G-ZIP, G-ZIGP) across six individual NME $C^{(e)}_{i,t}$ and combination $N_{i,t}$. Figure \ref{fig:nme-aic} reports AIC values across NME categories for the models; lower values indicate a better fit. We find that: \emph{(i) Poisson/ZIP are inadequate for several NMEs with long-tails, whereas grouping and generalized dispersion can deliver large gains.} For \textit{harsh\_braking} the G-ZIGP attains AIC 35{,}532 versus ZIP’s 114{,}153 and Poisson’s 150{,}901; for \textit{harsh\_acceleration} G-ZIGP yields 38{,}736 against G-ZIP’s 87{,}939 and ZIP’s 225{,}894. A similar pattern holds for \textit{too\_close\_distance}, where G-ZIGP yields 148{,}356 clearly improves upon ZIP/G-ZIP. These large AIC/BIC drops indicate that latent behavioral groups and non-Poisson dispersion are both needed to accommodate the extreme-count tail present in these NMEs. \emph{(ii) Portfolio-level performance favors Group-based models.} At the portfolio level, G-ZIGP also delivers the lowest AIC/BIC for the weekly total $N_{i,t}$ (309{,}980 / 310{,}166), followed by G-ZIP, with Poisson and ZIP far behind. Likelihood–based and error–based diagnostics offer a consistent yet nuanced perspective. From Poisson to ZIP, the total improves sharply in both deviance and point error (Poisson deviance mean $107.23$ to $48.83$, RMSE mean $143.46$ to $46.96$), reflecting ZIP’s ability to accommodate the mass at zero. Furthermore, G-ZIGP achieves the lowest deviance on the combination NME (with mean $32.17$) and the largest AIC/BIC gains on tail-prone components. Across secondary diagnostics (McFadden $R^2$, Brier, $\chi^2$), G-ZIGP excels precisely on the long-tailed components. 

In practice, a target-aware choice is recommended: use G-ZIGP for tail-prone NMEs (braking, acceleration, headway) to capture extreme-event drivers, and G-ZIP for NMEs closer to ZIP (such as serious speeding, forward collision, and lane departure). For portfolio-level dynamic ratemaking based on $N_{i,t}$, G-ZIP is the most reliable default, while component-level coaching and risk signaling benefit from G-ZIGP on the specific long-tailed dimensions.

\begin{table*}[htbp]
\centering
\scriptsize
\caption{Model evaluation metrics across NME categories.}
\label{tab:model_eval}
\setlength{\tabcolsep}{4pt}
\renewcommand{\arraystretch}{1.1}
\begin{tabular}{@{} l l r r r r r r r @{}}
\toprule
\multirow{2}{*}{\textbf{Model}} &
\multirow{2}{*}{\textbf{Metric}} &
\multicolumn{7}{c}{\textbf{NME categories}} \\
\cmidrule(lr){3-9}
& & \textbf{Harsh Braking} & \textbf{Harsh Accel.} & \textbf{Speeding Serious} & \textbf{Forward Collision} & \textbf{Lane Departure} & \textbf{Too Close Dist.} & \textbf{NME Total} \\
\midrule
\multirow[c]{9}{*}{\textbf{Poisson}}
  & AIC                     & 150901.36 & 436023.43 & 202800.43 & 189265.23 & 1360837.42 & 451410.99 & 1368398.72 \\
  & BIC                     & 150983.16 & 436105.22 & 202882.22 & 189347.03 & 1360919.21 & 451492.79 & 1368480.52 \\
  & Poisson deviance mean   & 13.90     & 44.25     & 14.92     & 12.01     & 110.81     & 33.47     & 107.23 \\
  & Poisson deviance std    & 10.02     & 45.65     & 3.82      & 0.80      & 18.85      & 5.29      & 36.86 \\
  & RMSE mean               & 21.70     & 46.95     & 16.21     & 14.53     & 97.04      & 39.17     & 143.46 \\
  & RMSE std                & 23.52     & 58.54     & 3.75      & 0.49      & 18.23      & 6.43      & 60.63 \\
  & Goodness-of-fit ($\chi^2$) & 706723.71 & 6156125.93 & 74942.68 & 35631.97 & 708894.07 & 101496.70 & 568803.97 \\
  & McFadden $R^2$ mean     & 0.31      & 0.34      & 0.10      & 0.29      & 0.10       & 0.14      & 0.09 \\
  & Brier zero mean         & 0.33      & 0.56      & 0.28      & 0.14      & 0.84       & 0.14      & 0.04 \\
\cmidrule(lr){2-9}
\multirow[c]{9}{*}{\textbf{ZIP}}
  & AIC                     & 114152.88 & 225893.61 & 167305.22 & 166058.97 & 170825.33 & 385343.83 & 1342582.83 \\
  & BIC                     & 114242.11 & 225982.84 & 167394.44 & 166148.20 & 170914.56 & 385433.06 & 1342672.06 \\
  & Poisson deviance mean   & 21.26     & 44.93     & 14.04     & 19.92     & 112.20     & 38.68     & 48.83 \\
  & Poisson deviance std    & 7.45      & 13.68     & 7.44      & 3.02      & 6.31       & 1.68      & 19.50 \\
  & RMSE mean               & 18.63     & 34.28     & 14.37     & 12.70     & 37.46      & 14.10     & 46.96 \\
  & RMSE std                & 28.36     & 59.90     & 6.31      & 2.33      & 18.69      & 18.81     & 62.98 \\
  & Goodness-of-fit ($\chi^2$) & 32833930.01 & 16933847.50 & 5892329.79 & 4946504.42 & 6575940.18 & 2460507.37 & 31833065.20 \\
  & McFadden $R^2$ mean     & 0.51      & 0.64      & 0.36      & 0.49      & 0.91       & 0.39      & 0.26 \\
  & Brier zero mean         & 0.25      & 0.14      & 0.26      & 0.35      & 0.27       & 0.34      & 0.36 \\
\cmidrule(lr){2-9}
\multirow[c]{9}{*}{\textbf{G-ZIP}}
  & AIC                     & 114178.88 &  87938.95 & 167331.22 & 166084.97 & 170851.33 & 385369.83 &  932321.42 \\
  & BIC                     & 114364.77 &  88124.84 & 167517.11 & 166270.86 & 171037.23 & 385555.73 &  932507.31 \\
  & Poisson deviance mean   & 6.44      & 35.52     & 14.11     & 11.97     & 113.99     & 29.24     & 78.69 \\
  & Poisson deviance std    & 5.74      & 26.10     & 4.21      & 0.81      & 20.70      & 7.16      & 28.58 \\
  & RMSE mean               & 20.84     & 53.01     & 15.90     & 14.50     & 97.00      & 36.90     & 129.81 \\
  & RMSE std                & 23.05     & 53.27     & 3.92      & 0.51      & 18.81      & 5.83      & 61.39 \\
  & Goodness-of-fit ($\chi^2$) & 195910.31 & 916818.57 & 75882.12 & 36261.41 & 713194.52 & 92428.31 & 332287.61 \\
  & McFadden $R^2$ mean     & 0.91      & 0.94      & 0.84      & 0.87      & 0.98       & 0.87      & 0.90 \\
  & Brier zero mean         & 0.20      & 0.12      & 0.21      & 0.12      & 0.11       & 0.12      & 0.04 \\
\cmidrule(lr){2-9}
\multirow[c]{9}{*}{\textbf{G-ZIGP}}
  & AIC                     & 35532.38  & 38736.38  & 74031.67  & 91595.21  & 57307.90  & 148355.53 & 309979.83 \\
  & BIC                     & 35718.27  & 38922.27  & 74217.56  & 91781.10  & 57493.79  & 148541.43 & 310165.72 \\
  & Poisson deviance mean   & 13.68     & 43.16     & 48.16     & 40.85     & 158.98     & 43.53     & 32.17 \\
  & Poisson deviance std    & 11.62     & 50.50     & 4.28      & 0.82      & 35.34      & 16.33     & 84.70 \\
  & RMSE mean               & 21.22     & 46.35     & 35.27     & 32.16     & 372.38     & 44.68     & 300.27 \\
  & RMSE std                & 24.19     & 59.11     & 2.00      & 0.26      & 21.58      & 12.85     & 30.89 \\
  & Goodness-of-fit ($\chi^2$) & 549464.63 & 3768498.38 & 73184.47 & 36592.24 & 752838.05 & 95478.28  & 749804.50 \\
  & McFadden $R^2$ mean     & 0.97      & 0.99      & 0.08      & 0.02      & 0.88       & 0.70      & 0.22 \\
  & Brier zero mean         & 0.20      & 0.12      & 0.25      & 0.18      & 0.25       & 0.17      & 0.21 \\
\bottomrule
\end{tabular}
\end{table*}

\subsection{Sensitivity Analysis}
\label{subsec:senana}

Table~\ref{tab:sa1} examines how the G-ZIP model reacts to the number of groups $G$ across six individual NME $C^{(e)}_{i,t}$ and combination $N_{i,t}$. We found that: \emph{(i) Monotone gains for the total with diminishing returns.} For $N_{i,t}$, AIC decreases from $1{,}342{,}583$ ($G{=}1$) to $735{,}690$ ($G{=}4$). BIC shows the same ordering. \emph{(ii) Heterogeneity is NME–specific.} Large benefits from grouping appear for most of the NME metrics. For example, \textit{harsh\_acceleration} (AIC $225{,}894$ to $37{,}895$ for $G{=}1$ to $4$), \textit{harsh\_braking} ($114{,}153$ to $35{,}887$ for $G{=}1$ to $4$). By contrast, \textit{too\_close\_distance} changes negligibly with $G$, implying that their variability is already well captured by ZIP’s zero–inflation and exposure terms rather than latent behavioral mixtures. \emph{(iii) Model selection guidance.} In Detail, the “elbow” typically occurs at $G{=}3$: gains from $G{=}3$ to $4$ are modest (e.g., \textit{harsh\_acceleration} $-9.4\%$, \textit{harsh\_braking} $-3.0\%$, \textit{lane\_departure} $-6.2\%$), so $G{=}3$ attains most of the improvement with better parsimony (favored by BIC). For $N_{i,t}$, both AIC and BIC continue to improve up to $G{=}4$, but the diminishing returns suggest choosing $G{=}3$ when interpretability and computational economy are prioritized, and $G{=}4$ when the best in-sample fit is required. Overall, results show that grouping is highly effective for NMEs exhibiting behavior like acceleration, braking, and lane departure, while others remain essentially homogeneous under ZIP.

\begin{table*}[htbp]
\centering
\small
\caption{Sensitivity Analysis on G-ZIP model: AIC/BIC across group numbers $G$ and NMEs.}
\label{tab:sa1}
\resizebox{\textwidth}{!}{
\begin{tabular}{@{} l c l r r r r r r r @{}}
\toprule
\multirow{2}{*}{\textbf{Model}} &
\multirow{2}{*}{\textbf{$G$}} &
\multirow{2}{*}{\textbf{Metric}} &
\multicolumn{6}{c}{$C^{(e)}_{i,t}$} &
\multicolumn{1}{c}{$N_{i,t}$} \\
\cmidrule(lr){4-9}\cmidrule(lr){10-10}
&&& \textbf{Harsh Braking} & \textbf{Harsh Accel.} & \textbf{Speeding Serious} & \textbf{Forward Collision} & \textbf{Lane Departure} & \textbf{Too Close Dist.} & \textbf{NME Total} \\
\midrule
\multirow[c]{8}{*}{\textbf{G-ZIP}}
  & \multirow{2}{*}{1} & AIC & 114152.88 & 225893.61 & 167305.22 & 166058.97 & 170825.33 & 385343.83 & 1342582.83 \\
  &                     & BIC & 114242.11 & 225982.84 & 167394.44 & 166148.20 & 170914.56 & 385433.06 & 1342672.06 \\
\cmidrule(lr){2-10}
  & \multirow{2}{*}{2} & AIC & 114178.88 &  87938.95 & 167331.22 & 166084.97 & 170851.33 & 385369.83 &  932321.42 \\
  &                     & BIC & 114364.77 &  88124.84 & 167517.11 & 166270.86 & 171037.23 & 385555.73 &  932507.31 \\
\cmidrule(lr){2-10}
  & \multirow{2}{*}{3} & AIC &  36987.35 &  41842.98 & 167357.22 & 166110.97 &  87900.90 & 385395.83 &  799843.67 \\
  &                     & BIC &  37269.91 &  42125.54 & 167639.77 & 166393.52 &  88183.46 & 385678.39 &  800126.23 \\
\cmidrule(lr){2-10}
  & \multirow{2}{*}{4} & AIC &  35886.62 &  37894.53 & 167383.22 & 166136.97 &  82433.76 & 385421.83 &  735690.44 \\
  &                     & BIC &  36265.85 &  38273.75 & 167762.44 & 166516.19 &  82812.98 & 385801.06 &  736069.66 \\
\bottomrule
\end{tabular}
}
\end{table*}

\begin{figure*}[htbp]
  \centering
 
  \begin{subfigure}[b]{0.25\textwidth}
    \centering
    \includegraphics[width=0.8\textwidth]{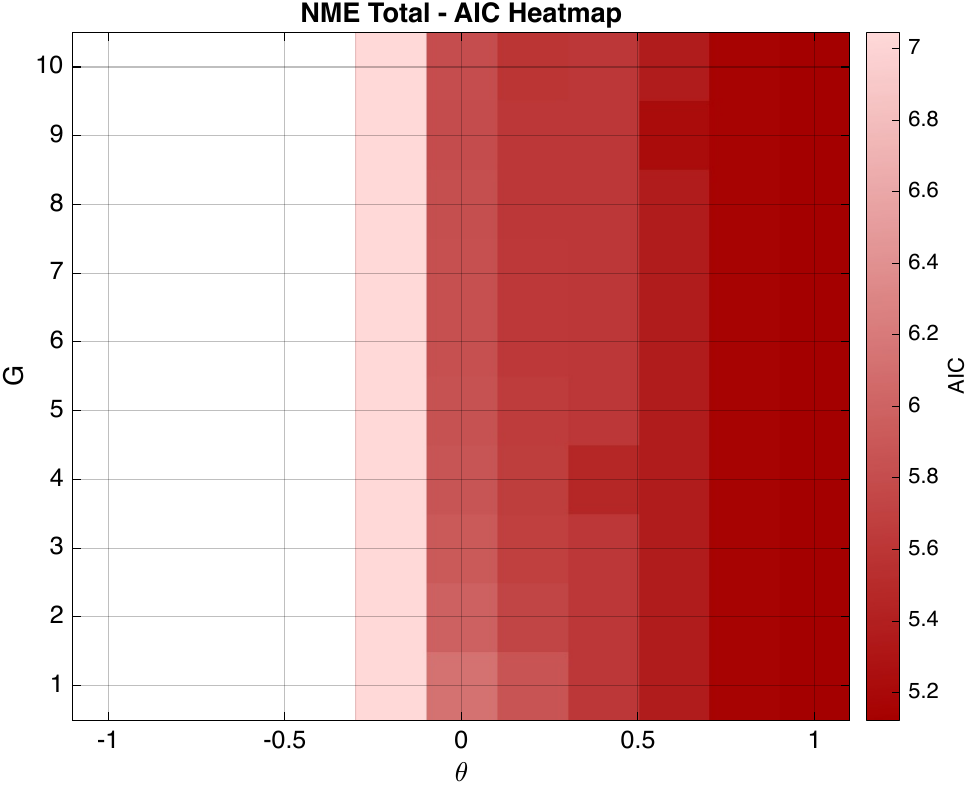}
    \label{fig:heat:1}
  \end{subfigure}\hfill
  \begin{subfigure}[b]{0.25\textwidth}
    \centering
    \includegraphics[width=0.8\textwidth]{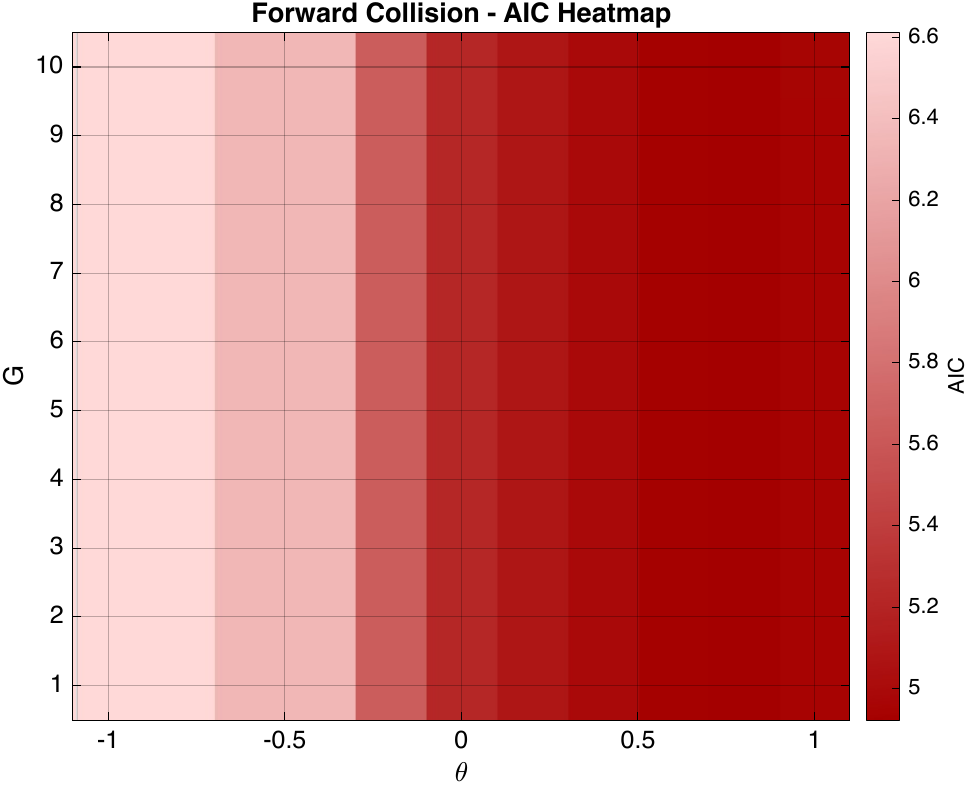}
    \label{fig:heat:2}
  \end{subfigure}\hfill
  \begin{subfigure}[b]{0.25\textwidth}
    \centering
    \includegraphics[width=0.8\textwidth]{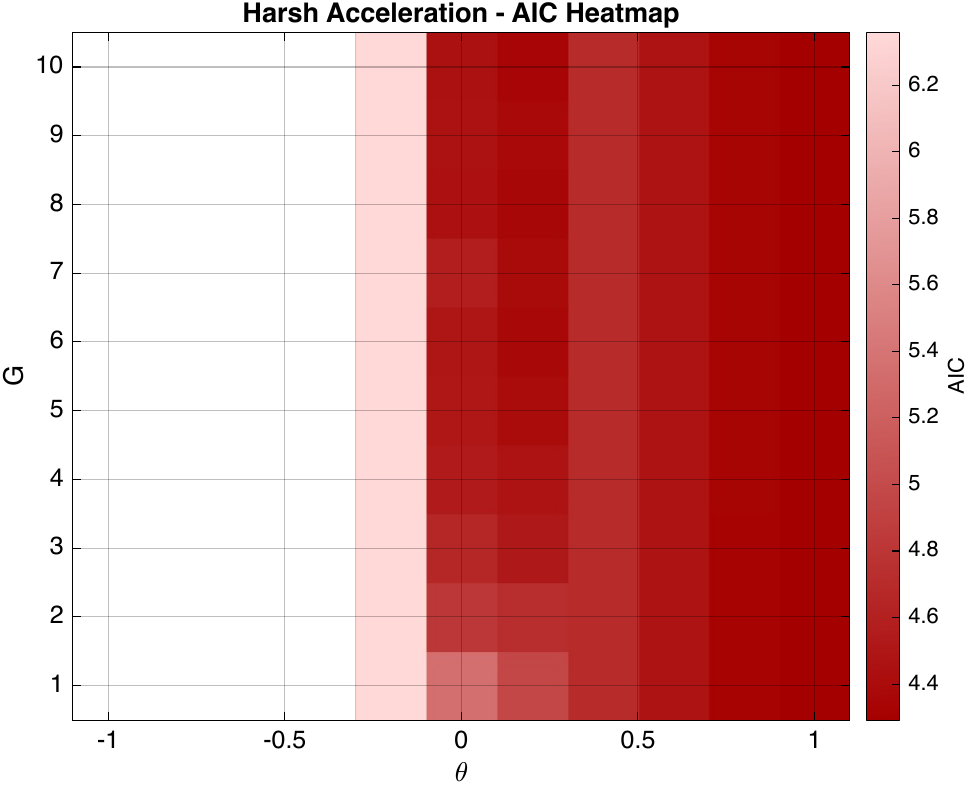}
    \label{fig:3}
  \end{subfigure}\hfill
  \begin{subfigure}[b]{0.25\textwidth}
    \centering
    \includegraphics[width=0.8\textwidth]{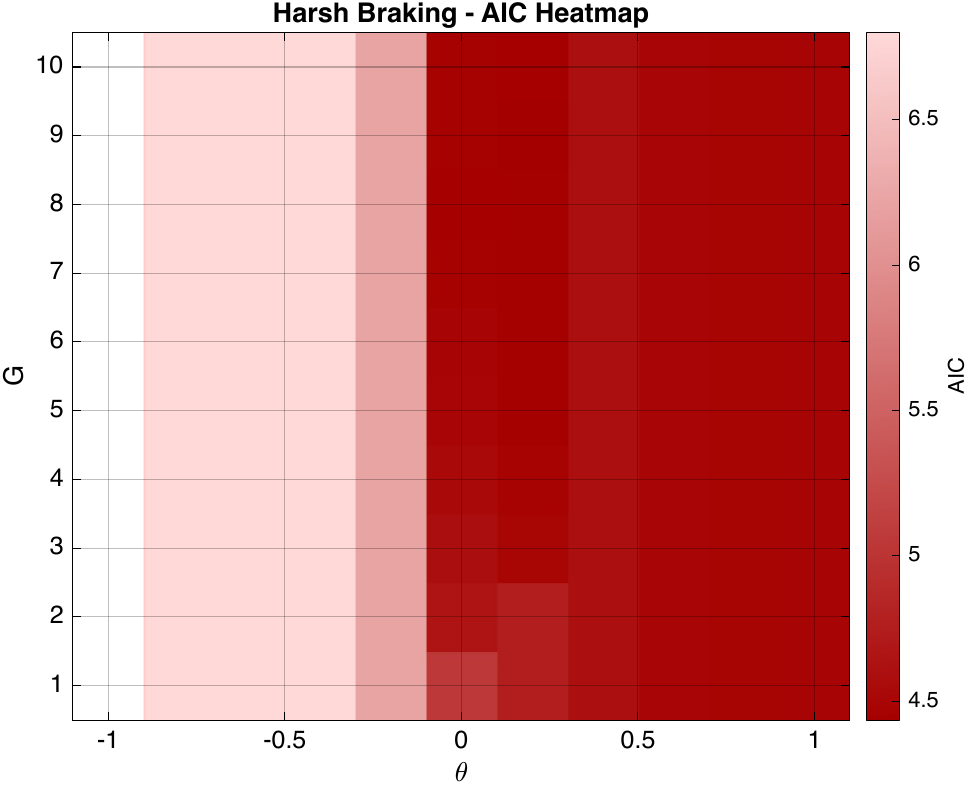}
    \label{fig:heat:4}
  \end{subfigure}

  \medskip %
 \hspace*{\fill}
  \begin{subfigure}[b]{0.25\textwidth}
    \centering
    \includegraphics[width=0.8\textwidth]{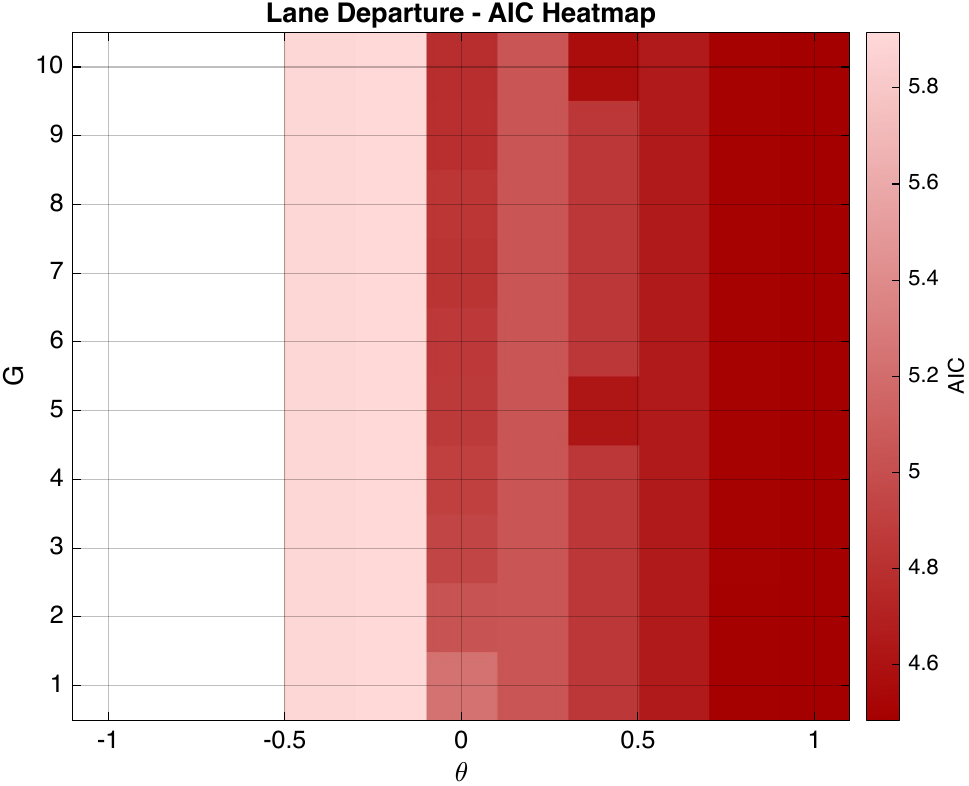}
    \label{fig:heat:5}
  \end{subfigure}\hfill
  \begin{subfigure}[b]{0.25\textwidth}
    \centering
    \includegraphics[width=0.8\textwidth]{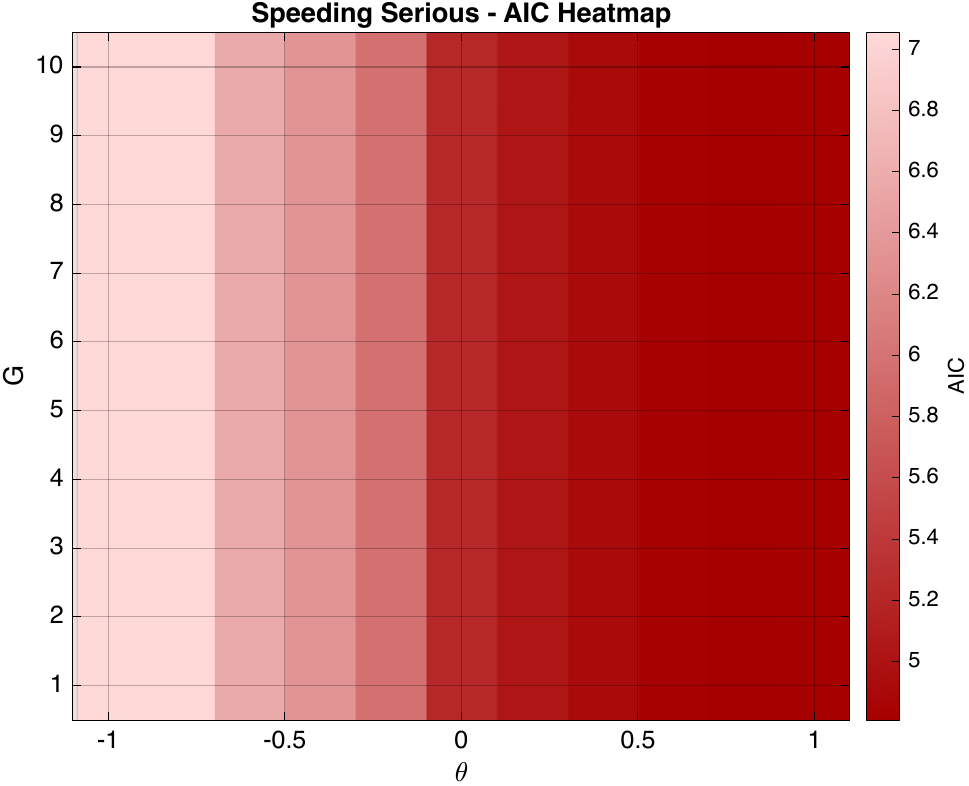}
    \label{fig:heat:6}
  \end{subfigure}\hfill
  \begin{subfigure}[b]{0.25\textwidth}
    \centering
    \includegraphics[width=0.8\textwidth]{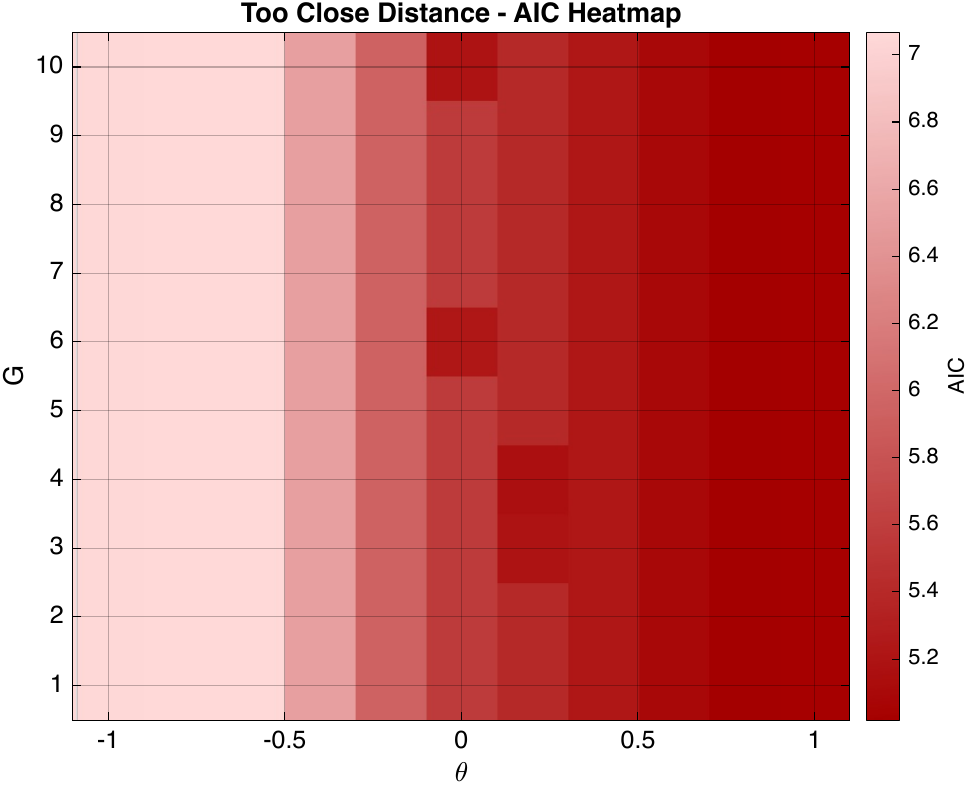}
    \label{fig:heat:7}
  \end{subfigure}
  \hspace*{\fill}
  \caption{Sensitivity analysis of AIC heatmaps across $G$ and $\theta$; darker red indicates lower AIC and white cells denote missing values.}
  \label{fig:aic_heatmaps}
\end{figure*}

We present a selected subset with $G\in\{1,2,3,4\}$ and generalized–Poisson dispersion $\theta\in\{-0.25,0,0.25,0.5\}$. Table~\ref{tab:sa2} reports AIC/BIC of the G-ZIGP under varying group numbers $G$ and generalized-Poisson dispersion $\theta$ across individual NME $C^{(e)}_{i,t}$ and the combination $N_{i,t}$. Figure \ref{fig:aic_heatmaps} shows the AIC performance across $G$ and $\theta$; darker red indicates a better fit (lower AIC). We found that: (i) \emph{Negative dispersion is strongly disfavored.} For $\theta=-0.25$, the performance deteriorates by orders of magnitude for the total (AIC $\approx 1.11\times 10^7$ for $G{=}1\!-\!4$), and similarly for several components, indicating that non-positive dispersion cannot explain the long-tails in our data. (ii) \emph{Moderate–high dispersion ($\theta\!\in\![0,0.5]$) is consistently beneficial, while the value of grouping is NME-dependent.} Take \textit{harsh\_acceleration} as an example, increasing $G$ yields large gains: AIC drops from $38{,}710$ ($G{=}1,\theta{=}0.5$) to $21{,}502$ ($G{=}4,\theta{=}0.5$), with the same ranking under BIC, evidencing meaningful latent heterogeneity. (iii) \emph{For the weekly total $N_{i,t}$, grouping helps when $\theta$ is small, whereas dispersion itself absorbs heterogeneity when $\theta$ is large.} At $\theta{=}0$, AIC/BIC decrease sharply as $G$ grows (AIC from $1{,}342{,}583$ to $755{,}976$ for $G{=}1$ to $4$). In contrast, at $\theta{=}0.5$ the best AIC/BIC are achieved by the model (AIC/BIC $309{,}954/310{,}043$ at $G{=}1$), with performance degrading slightly as $G$ increases. Overall, the evidence supports using $\theta{\ge}0$ throughout. When analyzing individual NME, $(G,\theta)=(4,0.5)$ attains the strongest fit. For $N_{i,t}$, two operating points emerge: a parsimonious yet best-scoring choice $(G,\theta)=(1,0.5)$, and a segmentation-friendly alternative around $\theta{=}0$ with $G{=}3$ to $4$ that substantially improves fit while enabling interpretable clusters.

\begin{table*}[htbp]
\centering
\small
\caption{Sensitivity Analysis on G-ZIGP model: AIC/BIC across group numbers $G$, $\theta$, and NMEs.}
\label{tab:sa2}
\resizebox{\textwidth}{!}{
\begin{tabular}{@{} l c c l r r r r r r r @{}}
\toprule
\multirow{2}{*}{\textbf{Model}} &
\multirow{2}{*}{\textbf{$G$}} &
\multirow{2}{*}{\textbf{$\theta$}} &
\multirow{2}{*}{\textbf{Metric}} &
\multicolumn{6}{c}{$C^{(e)}_{i,t}$} &
\multicolumn{1}{c}{$N_{i,t}$} \\
\cmidrule(lr){5-10}\cmidrule(lr){11-11}
&&&& \textbf{Harsh Braking} & \textbf{Harsh Accel.} & \textbf{Speeding Serious} & \textbf{Forward Collision} & \textbf{Lane Departure} & \textbf{Too Close Dist.} & \textbf{NME Total} \\
\midrule
\multirow[c]{32}{*}{\textbf{G-ZIGP}}
  & \multirow{8}{*}{1} & \multirow{2}{*}{-0.25} & AIC & 6320248.08 & 2283509.08 & 2163664.05 &  933899.74 & 810684.00  & 1828230.76 & 11076207.38 \\
  &                     &                         & BIC & 6320337.31 & 2283598.31 & 2163753.28 &  933988.97 & 810773.23  & 1828319.98 & 11076296.61 \\
\cmidrule(lr){3-11}
  &                     & \multirow{2}{*}{0}     & AIC & 114152.88  & 225893.61  & 167305.22  &  166058.97 & 170825.33  & 385343.83  & 1342582.83 \\
  &                     &                         & BIC & 114242.11  & 225982.84  & 167394.44  &  166148.20 & 170914.56  & 385433.06  & 1342672.06 \\
\cmidrule(lr){3-11}
  &                     & \multirow{2}{*}{0.25}  & AIC &  52243.15  &  80328.01  & 101934.24  &  117414.95 & 100566.13  & 236791.06  &  641338.82 \\
  &                     &                         & BIC &  52332.37  &  80417.24  & 102023.47  &  117504.18 & 100655.36  & 236880.29  &  641428.05 \\
\cmidrule(lr){3-11}
  &                     & \multirow{2}{*}{0.5}   & AIC &  35506.38  &  38710.38  &  74005.67  &   91569.21 &  57281.90  & 148329.53  &  309953.83 \\
  &                     &                         & BIC &  35595.61  &  38799.61  &  74094.90  &   91658.44 &  57371.13  & 148418.76  &  310043.06 \\
\cmidrule(lr){2-11}
  & \multirow{8}{*}{2} & \multirow{2}{*}{-0.25} & AIC & 6320274.08 & 2283535.08 & 2163690.04 &  933925.69 & 810710.00  & 1828256.74 & 11076233.38 \\
  &                     &                         & BIC & 6320459.97 & 2283720.97 & 2163875.93 &  934111.58 & 810895.89  & 1828442.64 & 11076419.27 \\
\cmidrule(lr){3-11}
  &                     & \multirow{2}{*}{0}     & AIC &  46075.15  &  67984.20  & 167331.22  &  166084.97 & 107958.33  & 385369.83  &  945896.15 \\
  &                     &                         & BIC &  46261.04  &  68170.09  & 167517.11  &  166270.86 & 108144.22  & 385555.73  &  946082.04 \\
\cmidrule(lr){3-11}
  &                     & \multirow{2}{*}{0.25}  & AIC &  52269.15  &  42523.27  & 101960.24  &  117440.95 & 100592.13  & 236817.06  &  486998.76 \\
  &                     &                         & BIC &  52455.04  &  42709.17  & 102146.14  &  117626.84 & 100778.03  & 237002.96  &  487184.65 \\
\cmidrule(lr){3-11}
  &                     & \multirow{2}{*}{0.5}   & AIC &  35532.38  &  38736.38  &  74031.67  &   91595.21 &  57307.90  & 148355.53  &  309979.83 \\
  &                     &                         & BIC &  35718.27  &  38922.27  &  74217.56  &   91781.10 &  57493.79  & 148541.43  &  310165.72 \\
\cmidrule(lr){2-11}
  & \multirow{8}{*}{3} & \multirow{2}{*}{-0.25} & AIC & 6320300.08 & 2283561.08 & 2163716.06 &  933951.74 & 810736.00  & 1828282.78 & 11076259.38 \\
  &                     &                         & BIC & 6320582.64 & 2283843.64 & 2163998.62 &  934234.30 & 811018.56  & 1828565.34 & 11076541.94 \\
\cmidrule(lr){3-11}
  &                     & \multirow{2}{*}{0}     & AIC &  39149.53  &  46711.21  & 167357.22  &  166110.97 &  87663.31  & 385395.83  &  817029.86 \\
  &                     &                         & BIC &  39432.09  &  46993.77  & 167639.77  &  166393.52 &  87945.87  & 385678.39  &  817312.42 \\
\cmidrule(lr){3-11}
  &                     & \multirow{2}{*}{0.25}  & AIC &  52295.15  &  33642.49  & 101986.24  &  117466.95 & 100618.13  & 236843.06  &  641390.82 \\
  &                     &                         & BIC &  52577.70  &  33925.05  & 102268.80  &  117749.51 & 100900.69  & 237125.62  &  641673.38 \\
\cmidrule(lr){3-11}
  &                     & \multirow{2}{*}{0.5}   & AIC &  35558.38  &  24595.03  &  74057.67  &   91621.21 &  57333.90  & 148381.53  &  310005.83 \\
  &                     &                         & BIC &  35840.94  &  24877.59  &  74340.23  &   91903.77 &  57616.46  & 148664.09  &  310288.39 \\
\cmidrule(lr){2-11}
  & \multirow{8}{*}{4} & \multirow{2}{*}{-0.25} & AIC & 6320326.08 & 2283587.08 & 2163742.05 &  933977.70 & 810762.00  & 1828308.75 & 11076285.38 \\
  &                     &                         & BIC & 6320705.30 & 2283966.30 & 2164121.27 &  934356.93 & 811141.22  & 1828687.97 & 11076664.60 \\
\cmidrule(lr){3-11}
  &                     & \multirow{2}{*}{0}     & AIC &  34665.73  &  35451.40  & 167383.22  &  166136.97 &  82119.48  & 385421.83  &  755976.43 \\
  &                     &                         & BIC &  35044.96  &  35830.62  & 167762.44  &  166516.19 &  82498.70  & 385801.06  &  756355.65 \\
\cmidrule(lr){3-11}
  &                     & \multirow{2}{*}{0.25}  & AIC &  52321.15  &  29032.52  & 102012.24  &  117492.95 & 100644.13  & 236869.06  &  641416.82 \\
  &                     &                         & BIC &  52700.37  &  29411.75  & 102391.47  &  117872.17 & 101023.36  & 237248.28  &  641796.04 \\
\cmidrule(lr){3-11}
  &                     & \multirow{2}{*}{0.5}   & AIC &  35584.38  &  21501.92  &  74083.67  &   91647.21 &  57359.90  & 148407.53  &  310031.83 \\
  &                     &                         & BIC &  35963.60  &  21881.14  &  74462.89  &   92026.43 &  57739.12  & 148786.75  &  310411.05 \\
\bottomrule
\end{tabular}
}
\end{table*}

\section{Conclusion}
\label{sec:conc}

This study demonstrates that integrating near-miss telematics into a group-based zero-inflated modeling framework substantially improves model fit compared to classical benchmarks. The proposed models capture both zero-excess and long-tail characteristics, enabling more accurate weekly prediction of risky driving behaviors. Future work includes exploring how external factors interact with driver behavior and near-miss risk. Explainable machine learning tools will enhance predictive performance and interpretability, allowing insurers to design personalized interventions and transparent premium adjustments.

\ifCLASSOPTIONcaptionsoff
  \newpage
\fi



%

\bibliographystyle{plain}
\bibliography{refs}

%








\end{document}